# High-Performance, High-Index-Contrast Chalcogenide Glass Photonics on Silicon and Unconventional Non-planar Substrates

Yi Zou[1], Danning Zhang[1,2], Hongtao Lin[1], Lan Li[1], Loise Moreel[1], Jie Zhou[3], Qingyang Du[1], Okechukwu Ogbuu[1], Sylvain Danto[4], J. David Musgraves[5], Kathleen Richardson[4], Kevin D. Dobson[6], Robert Birkmire[1,6], Juejun Hu[1,*]

[1] Department of Materials Science & Engineering, University of Delaware, Newark, Delaware, 19716, USA

[2] Center for Composite Materials, University of Delaware, Newark, Delaware, 19716, USA

[3] Department of Electrical & Computer Engineering, University of Delaware, Newark, Delaware, 19716, USA

[4] The College of Optics & Photonics, University of Central Florida, Orlando, Florida, 32816, USA

[5] IRradiance Glass, Inc. Orlando, Florida, 32816, USA

[6] Institute of Energy Conversion, University of Delaware, Newark DE, 19716, USA

*E-mail: hujuejun@udel.edu

**Abstract**

This paper reports a versatile, roll-to-roll and backend compatible technique for the fabrication of high-index-contrast photonic structures on both silicon and plastic substrates. The fabrication technique combines low-temperature chalcogenide glass film deposition and resist-free single-step thermal nanoimprint to process low-loss (1.6 dB/cm), sub-micron single-mode waveguides with a smooth surface finish using simple contact photolithography. Using this approach, the first chalcogenide glass micro-ring resonators are fabricated by thermal nanoimprint. The devices exhibit an ultra-high quality-factor of $4 \times 10^5$ near 1550 nm wavelength, which represents the highest value reported in chalcogenide glass micro-ring resonators. Furthermore, sub-micron nanoimprint of chalcogenide glass films on non-planar plastic substrates is demonstrated, which establishes the method as a facile route for monolithic fabrication of high-index-contrast devices on a wide array of unconventional substrates.

High-index-contrast (HIC) guided wave optical devices with a core-cladding index difference $\Delta n >$ 1 form the backbone of planar photonic integrated circuits since the strong optical confinement allows tight bends and small device footprints. The tight optical confinement and small mode volume in HIC devices further leads to significant enhancement of photon-matter interactions, which underlies their roles as on-chip light sources,[1] photonic sensors,[2] and nonlinear optical devices.[3] In addition to planar photonic applications, HIC optical structures also constitute key elements in diffractive optics, micro-lens arrays with low spherical aberration, and broadband grating reflectors.[4, 5, 6] Despite these apparent advantages, microscopic defects and sidewall roughness inevitably induced by standard lithography and plasma etching processing result in large scattering losses in HIC structures, since scattering loss scales with the square of index contrast.[7, 8] Fine-line lithography tools such as Deep-UV lithography and electron beam lithography have thus been regarded as a must for fabrication of low-loss HIC devices given their small feature size and high sensitivity to sidewall roughness. Additionally, to unleash the potential of HIC devices, a number of methods were applied to different material systems to smooth the device surface and reduce optical losses such as oxidation smoothing[9], post-fabrication wet etching[10], hydrogen annealing in $Si/SiO_2$ systems[11], laser-reformation[12], and thermal reflow of polymer[13] and glass materials.[14] However, most of these methods involve high temperature and multiple processing steps, which pose a limitation for efficient process methodologies and a challenge on device integration.

Here we present a new method to create low-loss HIC photonic devices using simple, low-cost UV lithography on a standard contact mask aligner. The novelty of our approach is two-fold: firstly, we capitalize on the low deposition temperature and amorphous nature of compositionally-engineered, high-refractive-index ChG alloys to enable monolithic fabrication of HIC photonic devices on a variety of substrate materials. ChGs are ideal candidates for an array of photonic applications because of their wide infrared transparency window,[15, 16] large Kerr nonlinearity,[17, 18] high refractive indices,[19, 20] tailorable photosensitivity,[21, 22] and large photothermal figure-of-merit.[23, 24] Thermal evaporation has been successfully employed to derive high-quality ChG films over large substrate areas at temperature < 50 ℃. Given the low processing temperature and the demonstrated direct patterning capability on plastic substrates, our technique can potentially be fully compatible with CMOS backend integration as well as roll-to-roll processing.

Secondly, we synergistically combined thermal reflow modification of resist pre-forms and single-step nanoimprint to create sub-micron, single-mode HIC photonic devices with record optical performance. Thermal nanoimprint has been recognized as an emerging technology that promises low cost and high-throughput patterning of micro- and nano-scale structures [25, 26, 27] and has the potential for roll-to-roll continuous patterning.[28, 29, 30] Since nanoimprint lithography is based on mechanical molding of materials, which is quite different from traditional lithographic techniques, the materials used in imprinting need to be easily deformable under applied pressure at elevated temperatures.[31] Thus, conventional crystalline semiconductor materials, such as silicon, are not suitable to be used as imprint resist materials. Therefore, nanoimprint processing of silicon and other high-index semiconductors requires an extra etching step to transfer patterns from imprint resists to the device layer. Several groups have demonstrated the patterning of polymer photonic devices by direct imprint methods,[32, 33, 34, 35, 36, 37] but polymers lack the aforementioned advantages associated with HIC devices. ChGs, on the other hand, have suitable softening characteristics due to their amorphous nature,[38] and they can be directly molded to desired shapes without resorting to extra pattern transfer steps. Thermal nanoimprint of ChG films has been used to pattern large-core waveguides,[39, 40, 41] diffraction gratings[42, 43, 44] and wire-grid polarizers;[45] however, imprint fabrication of single-mode, sub-micron HIC devices (e.g. waveguides and resonators) has not been demonstrated. In addition, while resist pre-form reflow was previously adopted to produce microlenses and multi-level structures,[46, 47, 48] here we present the first effort to exploit the method's unique ability to produce a smooth surface finish required for low-loss photonic device processing.

The following sections are organized as follows: we start with examining the basic material properties, the rationales for glass composition selection, as well as the underlying physics that governs the kinetics of geometry shaping in the imprint process. Kinetic theory and experimental results presented below confirms that the smooth stamp surfaces effectively translate to a high-quality imprinted surface finish and, hence, low-loss optical performance. We subsequently validate the theoretical insight through optical characterization of HIC glass resonator devices fabricated on rigid semiconductor substrates using the single-step imprint process. Lastly, we report the direct imprint patterning of photonic devices on flexible polymer substrates, which clearly establishes the method as a straightforward processing route for monolithic integration of HIC devices on a wide array of unconventional substrates.

**Results**

**1. Nanoimprint Process Flow**

In **Figure 1**, we depict the process flow to pattern HIC ChG photonic devices by nanoimprint. Photoresist line patterns were first defined on a pristine silicon substrate followed by thermal reflow of the photoresist lines to smooth their sidewalls. The surface cleanliness of Si wafers used in the process is crucial, as any contaminants or particulates can result in pinning of the resist line edges.[49] A polydimethylsiloxane (PDMS) elastomer stamp was subsequently replica molded from the resist line patterns. High-quality ChG films were deposited using vacuum thermal evaporation (refer to Methods section for details of the film deposition process). Imprint on ChG films was then performed by loading the PDMS stamp onto the films under pressure at elevated temperatures. In the last fabrication step, a 4 μm-thick SU-8 epoxy layer was spin coated on top of the patterned devices to serve as a top cladding to protect the devices from surface oxidation.

## 2. As-Se Chalcogenide Glass Film Characterizations

Two glass compositions in the binary As-Se system were selected for imprint experiments: $As_{40}Se_{60}$ (i.e. stoichiometric $As_2Se_3$), and $As_{20}Se_{80}$. The rationales for these composition choices will be detailed in the succeeding sections. **Figure 2** a shows a cross-sectional SEM image of as-deposited $As_{20}Se_{80}$ films, showing a uniform thickness with dense micro-structures free of defects. As shown in the dispersion diagrams measured using ellipsometry (Figure 2 b), the refractive indices of as-deposited $As_{40}Se_{60}$ and $As_{20}Se_{80}$ films at 1550 nm wavelength were 2.72 and 2.61, respectively, sufficiently high for compact on-chip integration. Both films are transparent in the telecommunication wavelength.

Micro-Raman spectroscopic characterization was carried out on as-deposited and thermally imprinted $As_{20}Se_{80}$ films (Figure 2 c) to examine structural modifications induced by the thermal imprint process. The dominant feature in the glass Raman spectra is the broad band located at 110 – 300 $cm^{-1}$. The band can be further deconvolved into six peaks, each of which corresponds to a distinctive vibrational mode associated with a specific structural group. In order to understand the structure variation, the spectra were fitted while keeping the position of the peaks constant.[50, 51]

Figure 2 d illustrates the fitting results for as-deposited and imprinted $As_{20}Se_{80}$ films. Based on references [51, 52], peaks located at 125 cm$^{-1}$, 212 cm$^{-1}$ 223 cm$^{-1}$, 241 cm$^{-1}$, 257 cm$^{-1}$, and 269 cm$^{-1}$ are assigned to the vibrational modes of As-As units, $As_4Se_4$ units, $AsSe_3$ pyramidal units, Se-Se chain, Se-Se ring and Se-Se bridge, respectively (**Table 1**). We plot the area of each fitted peak over the total area of all peaks combined in Figure 2 e. After thermal imprint, the fractions of $AsSe_3$ pyramidal unit and Se-Se bridge unit increase, which clearly indicates an increased degree of polymerization and cross-linking in the glass network. This observation is also consistent with previous ChG film annealing studies.[53] Compared to the stoichiometric $As_{40}Se_{60}$ film, Raman spectra from $As_{20}Se_{80}$ films show reduced fraction of the $AsSe_3$ unit, which is replaced by Se-Se homopolar bonds and corresponds to a decreased glass network connectivity. No sharp Raman peaks were observed after imprint, which suggests that the glass structure remained amorphous. The amorphous structure of $As_{40}Se_{60}$ and $As_{20}Se_{80}$ thin films after thermal imprint is further confirmed by X-ray diffraction (XRD) analysis. Both films exhibited broad, featureless XRD spectra without noticeable peaks (not shown here), which confirms the absence of detectable crystallization in the films after thermal nanoimprint. This result is critical to low-loss optical transmission in the imprinted devices.

### 3. Kinetics of Geometry Shaping in Nanoimprint

Both previous reports[31, 54] and our own experimental observations (Supporting Information Part I) suggest that the nanoimprint process is kinetically controlled by the viscous flow of glass. Therefore, the two key experimental parameters that dictate the imprint quality are glass viscosity at the imprint temperature and the stamp loading pressure. For a given film thickness and stamp geometry, simple dimensional analysis shows that the deformation rate of glass during imprint scales with the factor $P/\eta$, where P is the loading pressure (in Pascal) and $\eta$ denotes the dynamic viscosity of glass at the imprint temperature (in Pascal second, Pa s). The viscosity of glass strongly depends on temperature as described by the Vogel-Fulcher-Tamann (VFT) model[55]:

$$\log_{10} \eta = A + \frac{B}{T-C} \qquad (1)$$

where T is the temperature (in Kelvin), and A, B, and C are material-specific constants determined by the glass network configuration. The two glass compositions we selected, $As_{40}Se_{60}$ and $As_{20}Se_{80}$, possess mean coordination numbers $\langle r \rangle$ of 2.4 and 2.2, respectively. They represent a structural evolution from a fully interconnected network ($As_{40}Se_{60}$) of $AsSe_3$ heteropolar structural units bound to form a puckered layer structure, to a cross-linked chain-like network ($As_{20}Se_{80}$) comprised of primarily homopolar Se-Se bonded chains cross-linked by occasional $AsSe_3$ pyramidal units.[56] Raman spectra shown in Figure 1 clearly indicate the presence of homopolar Se-Se bonds in the $As_{20}Se_{80}$ composition. Based on measured variation in viscosity with arsenic content and temperature, we anticipate a large increase of glass viscosity at the glass transition temperature ($T_g$) as the arsenic content increases in our deposited films. This trend, measured for bulk materials by Musgraves *et al.*,[38] quantified this variation, where the constants A, B, and C for these glasses were experimentally evaluated. The A, B and C parameters we used for glass viscosity calculations are tabulated in Supporting Information Part I, and temperature-dependent viscosity of the two binary glasses is plotted in **Figure 3** a, where the x- axis is referenced with respect to $T_g$ of the two compositions (382 K and 466 K for $As_{20}Se_{80}$ and $As_{40}Se_{60}$, respectively). Given the strong dependence of viscosity on temperature (Eq. 1), optimization of the imprint process was carried out by varying the imprint temperature at a fixed loading pressure (0.13 MPa). The significant impact of temperature on imprinted device geometry is clearly visible from Figure 3 b, which shows the cross-sectional profiles of single-mode ridge waveguides imprinted on $As_{20}Se_{80}$ films at different temperatures. The waveguide morphology evolves from a shallow ridge to complete trench filling as the imprint temperature rises from 160 °C to 166 °C. The observed morphology change was quantitatively described using fluid dynamics finite element simulations using glass viscosity values calculated from Eq. 1. (Supporting Information Part I) and the simulated waveguide profiles are shown in Figure 3 c for comparison. The fluid dynamics processing modeling enables quantitative, a priori, prediction of imprinted waveguide ridge height, as proven in Figure 3 d.

The fundamental understanding of the imprint kinetics allows us to quantitatively analyze the line edge roughness evolution in the molding process, a critical step for the reduction of optical loss. The line edge roughness values of the reflowed resist line template, PDMS stamp replica molded from the template, as well as the final imprinted waveguide structure were 0.9 nm $\pm$ 0.3 nm, 0.9 nm $\pm$ 0.2 nm, and 0.9 nm $\pm$ 0.2 nm, respectively (**Figure S2** in Supporting Information Part II), measured using atomic force microscopy (AFM). Despite the soft and easily deformable nature of the PDMS stamp, we see that the elastomer imprint process is capable of reproducing the ultra-smooth surface finish of the resist template structure. This observation was modeled by solving the roughness evolution kinetics in the imprint process (Supporting Information Part II). The roughness amplitude exponentially decays during imprint with a decay time constant given by:

$$\tau = \frac{(1-v-2v^2)^2}{1-v} \times \frac{\eta}{E} \tag{2}$$

where $v$ and $E$ denotes the Poisson ratio and Young's modulus of the PDMS stamp, respectively. Eq. 2 gives a typical roughness decay time constant < 1 s. The short time scale indicates that elastomer imprint is capable of generating an atomically smooth surface finish, provided that the reflow modification of resist template effectively removes any residue roughness. In addition, compared to post-fabrication thermal reflow treatment on patterned glass waveguides,[14] our new approach circumvents the trade-off between roughness reduction and optical property degradation of chalcogenide glass materials by eliminating parasitic optical losses associated with glass crystallization or partial vaporization, since the reflow heat treatment is performed on a resist template rather than directly on the glass devices. Both XRD and Raman measurements confirmed that the $As_{20}Se_{80}$ film maintained its amorphous structure after thermal imprint. The main structural modification induced by the imprint process was an increased degree of polymerization evidenced by micro-Raman analysis, which was an anticipated result of the heat treatment and an enhancement of the cross-linking due to the relaxation/reformation of defective bonds presented in the as-deposited film network.[20, 57]

## 4. Optical Characterizations of Imprinted Devices

Since viscosity is the critical parameter that kinetically controls the imprint process, the optimal imprint conditions vary significantly across the selected glass compositions given their drastically different viscosity-temperature behaviors. The optimal temperature and time for imprinting thermally evaporated $As_{20}Se_{80}$ films are set at 166 ℃ for 15 minutes, optimized based on SEM observations of imprinted device morphology and optical loss measurements. In comparison, thermally evaporated $As_{40}Se_{60}$ films were imprinted at a dwell temperature of 270 ℃ due to the composition's much higher viscosity. Since the dwell temperature is close to the denaturing temperature of PDMS, imprinted devices with the best optical performances (i.e. lowest optical loss) were attained in thermally evaporated $As_{20}Se_{80}$ films. The detailed imprint conditions and resulting device optical performances have been summarized in Supporting Information Part IV.

Notably, despite their vastly different viscosity-temperature behavior, nanoimprint fabrication in $As_{20}Se_{80}$ and $As_{40}Se_{60}$ were performed at similar viscosity values, which further corroborates that the imprint kinetics is governed by viscous flow in the glass materials.

**Figure 4** a and b are optical microscope images showing pulley-coupled micro-rings resonators imprinted on thermally evaporated $As_{20}Se_{80}$ films deposited on silicon wafers with 3 µm thermal oxide. Instead of using a simple "groove" on the PDMS stamp to define a ridge waveguide structure, we imprinted a pair of isolation trenches on both sides of the waveguide core, designed to displace the minimal amount of glass possible. The "minimum fluid displacement" strategy pioneered by Luther-Davies *et al.* had been successfully applied to both polymer and glass imprints.[39, 58] The morphology of the coupling region between the bus waveguide and the micro-ring has been characterized by AFM (Figure 4 c) confirming excellent pattern fidelity and a smooth surface finish.

Figure 4 d plots a typical measured TE-polarization transmission spectrum of a ring resonator imprinted in thermally evaporated $As_{20}Se_{80}$ films. By systematically varying the gap width between the bus waveguide and the micro-ring, the resonator was designed to operate in the over coupling regime near 1550 nm wavelength. Cavity quality factors (Q), defined as the ratio of wavelength against the resonant peak full width at half maximum, were measured by averaging over multiple devices. The measurement yielded an average loaded Q-factor of 100,000 in imprinted $As_{20}Se_{80}$ devices. A maximum loaded Q-factor as high as 150,000 was observed (Figure 4 e). Such a high Q-factor corresponds to an equivalent waveguide loss of 1.6 dB/cm and an intrinsic Q-factor of 390,000, the highest value reported in ChG micro-ring resonators. It is worth noting that our present result was obtained on devices patterned using a simple contact aligner, yet the Q value represents a 20-fold improvement over the Q's of our previously demonstrated ChG micro-ring resonators fabricated on an i-line stepper.[59]

Optical loss in a micro-ring resonator can be attributed to several mechanisms. Specifically, the total loss is written as:

$$\alpha_{tot} = \alpha_{abs} + \alpha_{sca} + \alpha_{bend} + \alpha_{sub} \qquad (3)$$

where $\alpha_{abs}$, $\alpha_{sca}$, $\alpha_{bend}$, and $\alpha_{sub}$ denote optical losses associated with material absorption, surface and sidewall roughness scattering, waveguide bending, and substrate leakage, respectively. To assess the loss mechanisms in the imprinted micro-ring devices, we performed finite difference modal simulations using a full-vectorial bending mode solver incorporating cylindrical perfectly matching boundary layers (CPML).[60] The model used waveguide dimensions experimentally measured from cross-sectional SEM images. According to the simulation results, substrate leakage loss is negligible and radiative leakage due to waveguide bending causes 0.1 dB/cm loss in our micro-ring device. The roughness scattering loss was estimated using the Payne-Lacey theory and surface roughness measured by AFM to be < 0.1 dB/cm.[61] Therefore, we conclude that material attenuation due to the $As_{20}Se_{80}$ glass and the SU-8 cladding layer accounts for approximately 1.5 dB/cm optical loss. This material loss figure is consistent with our cut-back measurement of optical loss in large-core imprinted As-Se glass waveguides, where material contribution dominates over other loss mechanisms. The material loss can be mitigated through purification of raw bulk glass materials as well as further processing optimization. Therefore, we expect that our single-step imprint fabrication technique will likely lead to HIC single-mode photonic devices with optical loss well below 1 dB/cm.

## 5. Direct Monolithic Imprint Fabrication on Nonplanar Substrates

We further demonstrated that the nanoimprint method is also applicable to HIC photonic integration on soft polymer substrates, for example polyimide and polyethylene terephthalate (PET) plastic substrates, given the amorphous nature and low deposition temperature of ChG films. Details of the plastic substrate imprint process are elaborated in the Experimental Section. **Figure 5** a is a macroscopic view of the imprinted $As_{20}Se_{80}$ glass devices on a 70 μm thick PET film, Figure 5 b shows an optical microscope image of the devices, and Figure 5 c plots the measured transmission spectrum of a flexible micro-ring resonator. The flexible resonators exhibited an average Q value of 10,000. The number is much lower compared to devices printed on Si substrates but comparable to previous reports on optical resonator structures fabricated on flexible substrates.[62] The lower Q is likely due to the inferior surface quality of the polyimide substrate. Compared to the standard transfer printing protocols for device fabrication on flexible substrates, the direct imprint patterning approach offers a monolithic integration alternative with potentially improved throughput and yield, and may also enables roll-to-roll processing of HIC photonic devices over large substrate areas inaccessible using conventional lithographic patterning methods.

## Discussion

To conclude, we report a simple, substrate-blind direct nanoimprint approach for high-performance, high-index-contrast glass photonic device fabrication. The method combines resist reflow and thermal nanoimprint to achieve an ultra-smooth surface finish on fabricated devices and hence low optical loss. We used the technique to demonstrate ChG micro-ring resonators with a record intrinsic Q-factor of 390,000 at 1550 nm wavelength. Furthermore, the low processing temperature of our technique is fully compatible with photonic integration on flexible polymer substrates, and we successfully applied the method to monolithically fabricate HIC glass photonic devices on plastic substrates. The technique is also compatible with CMOS backend integration given its minimal thermal budget requirement, making it potentially attractive for an array of emerging applications ranging from optical interconnects to conformal photonic sensor integration on complex curvilinear surfaces.

**Methods**

Bulk $As_{40}Se_{60}$ and $As_{20}Se_{80}$ glasses were prepared using a traditional melt-quenching technique.[38] From the bulk glass target material, 360 nm-thick thin films were deposited by thermal evaporation. Thermal evaporation of powdered glass was carried out at a base pressure $< 10^{-6}$ Torr in a custom-designed single-source evaporator (PVD Products, Inc.). The deposition substrates are 3" silicon wafers coating with 3 μm thick thermal $SiO_2$. The deposition rate was maintained at ~ 16 Å/s. The substrate upon which the film was deposited was maintained at room temperature throughout the deposition. Thermal evaporation deposition conditions were optimized with respect to the optical performance of the deposited films (e.g. thickness uniformity and surface quality) and were extremely repeatable. The substrates used for thin film deposition include 6" Si wafers with 3 μm thermal oxide (Silicon Quest International Inc.), 30 μm-thick Polyimide film (DuPont Inc.), and 80 μm-thick PET films (GoodFellow Inc.). More details concerning the bulk glass preparation and film deposition process may be found elsewhere [63, 64].

An NR9-1000PY photoresist (Futurrex inc.) pattern was first defined on a Si wafer using contact photolithography on an Karl Suss MJB-3 mask aligner, and then a thermal treatment at 135 ℃ for 5 s was implemented to reflow the NR9 polymer resist and create a smooth surface finish. This photoresist pattern was used as the master mold which can be conveniently used many times (> 20 times in our experiments) as the template to produce PDMS soft stamps. 5 mm thick elastomer stamps were made by casting liquid PDMS (Sylgard 184, Dow Corning Inc., 5:1 mixing ratio between the monomer and the curing agent) onto the master mold, first baked at 80 ℃ for 12 hours and then baked at 110 ℃ for 5 hours to ensure that the PDMS was fully cured.

Imprint was performed in a glove box purged by $N_2$ to protect the chalcogenide glass films from oxidation. The glass film sample along with the soft stamp were placed on a hotplate pre-set at imprint temperature. Imprint pressure of approximately 0.13 MPa was applied by loading a metal block on the film-stamp assembly. After imprint, the hotplate was allowed to cool to below 60 ℃ at a ramp-down rate of 5 ℃/min, before the sample was removed from the hotplate. The stamp was then manually delaminated from the glass film sample.

Refractive index dispersion of the ChG films was measured using a M-44 variable angle spectroscopic ellipsometer (J.A. Woollam Co., Inc.). Ellipsometry data were collected at three incidence angles: 68 °, 73 °, and 78 °. This interval covers the (pseudo) Brewster angle for the chalcogenide glass film samples.

The micro-Raman spectra for the as-deposited and imprinted films were recorded using a Bruker Senterra Raman spectrometer with a Raman microprobe attachment. This system has a typical resolution of 2-3 cm$^{-1}$ at room temperature and uses a backscattering geometry. The system consists of an edge filter for Rayleigh rejection, a microscope equipped with $\times 10$, $\times 50$ and $\times 100$ objectives and a CCD detector. A 785 nm NIR semiconductor laser was used for excitation with an incident power of approximately 2 mW. The use of a 785 nm source with a low power was specific to our study in order to avoid any photo-structural changes which the laser beam might induce in the samples during measurement. Three measurements were performed on each sample to ensure reproducibility of the spectra fitting.

The surface morphology and roughness was measured using tapping mode atomic force microscopy (AFM) on a Dimension 3100 (Digital Instruments, Inc.) microscope. To accurately measure the waveguide line edge roughness, AFM line scans were performed parallel to the waveguides to avoid line-to-line variations in a 2-D AFM scan. The scans were performed at 5 or more different locations on each sample and the results were averaged. Silicon AFM probes (Tap 150-G from Budget Sensors, Inc) with a force constant of 5 N/m and a resonant frequency of 150 KHz were used. The cross sectional images of imprinted waveguides/resonators were taken on a JSM-7400F (JEOL, Inc.) scanning electron microscope (SEM).

Before optical measurements, a 3 μm-thick SU-8 polymer (MicroChem Inc.) layer was spin coated on the patterned glass samples, which served as a top cladding and also prevented the glass films from oxidation. The waveguide facets on flexible substrates were prepared using focused $Ga^{2+}$ ion beam milling (beam current 4 nA, accelerating voltage 30 kV) using a Zeiss Auriga 60 CrossBeam$^{TM}$ FIB nanoprototyping workstation. The transmission spectra of the micro-ring resonators were measured using a tunable laser (Agilent Technologies Model 81682A) operating in a step sweep mode. We used a fiber end-fire coupling method for coupling light into and out of the devices. All the measurement results reported in this paper use TE-polarized light.

**Acknowledgements**

The authors would like to thank funding support provided by the Department of Energy under award number DE-EE0005327 and the University of Delaware Research Foundation (UDRF). UCF co-authors acknowledge funding provided in part by the US Department of Energy [Contract # DE- NA000421], NNSA/DNN R&D.

**Disclaimer**

This paper has been prepared as an account of work partially supported by an agency of the United States Government. Neither the United States Government nor any agency thereof, nor any of their employees, makes any warranty, express or implied, or assumes any legal liability or responsibility for the accuracy, completeness or usefulness of any information, apparatus, product or process disclosed, or represents that its use would not infringe privately owned rights. Reference herein to any specific commercial product, process, or service by trade name, trademark, manufacturer, or otherwise does not necessarily constitute or imply its endorsement, recommendation, or favoring by the United States Government or any agency thereof. The views and opinions of authors expressed herein do not necessarily state or reflect those of the United States Government or any agency thereof.

**Author Contributions**

Y.Z., L.M. and J.Z. conducted optical modeling, material characterizations, device fabrication and testing. D.Z. performed imprint modeling and analysis. H.L and L.L. assisted in master mold fabrication. K.D. assisted in material characterizations. J. H. conceived the device and structural designs. S. D., J. D. M., and K. R. contributed to material synthesis. J. H., K. R. and B.B. supervised and coordinated the project. All authors contributed to writing the paper.

**Competing financial interests**

The authors declare no competing financial interests.

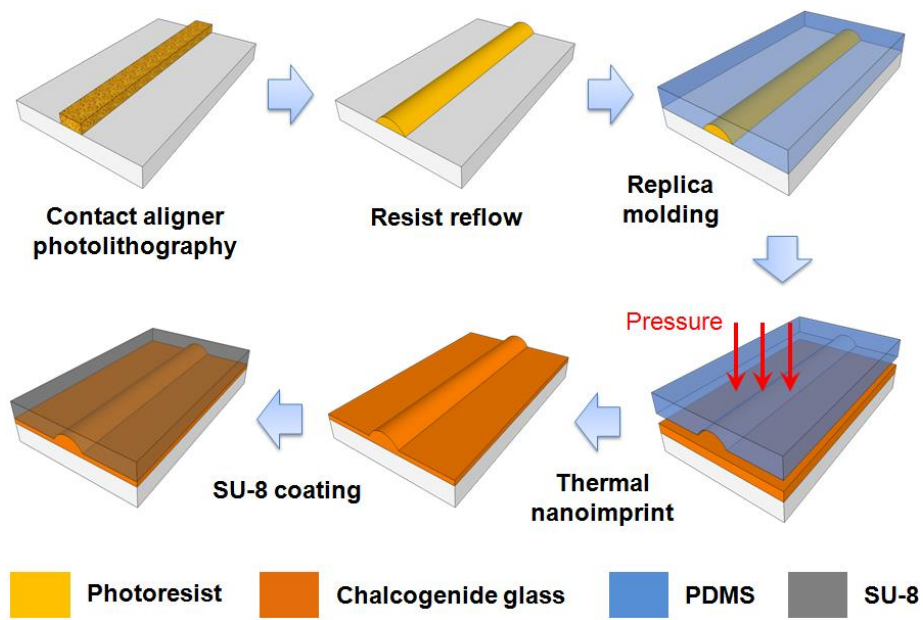

**Figure 1.** Schematic illustration of the thermal nanoimprint fabrication process.

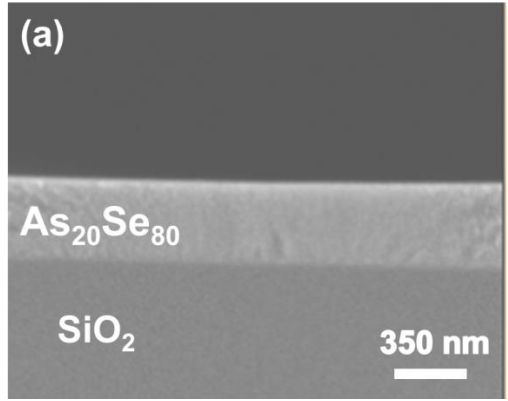

(a)

As$_{20}$Se$_{80}$

SiO$_2$

350 nm

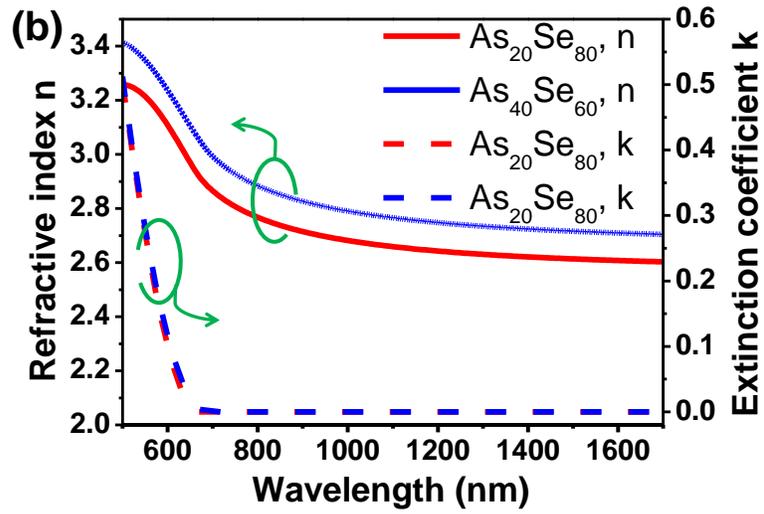

(b)

As$_{20}$Se$_{80}$, n
As$_{40}$Se$_{60}$, n
As$_{20}$Se$_{80}$, k
As$_{20}$Se$_{80}$, k

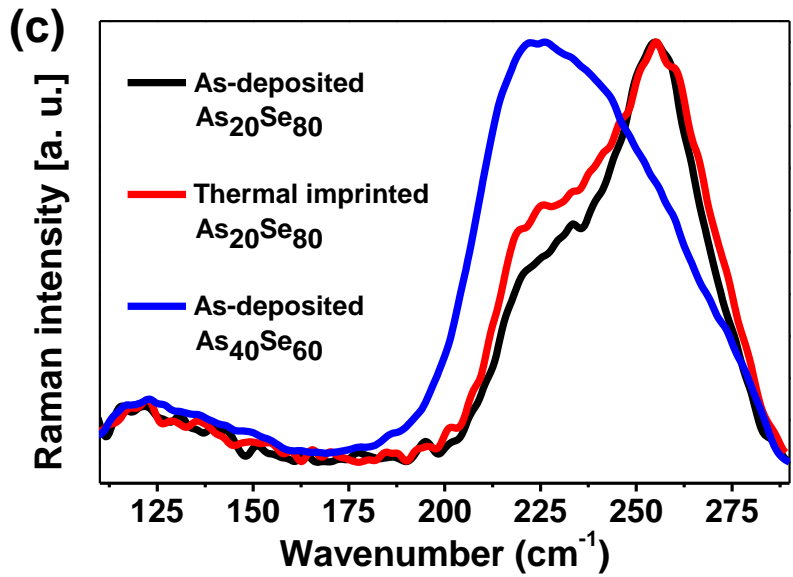

(c)

As-deposited
As$_{20}$Se$_{80}$

Thermal imprinted
As$_{20}$Se$_{80}$

As-deposited
As$_{40}$Se$_{60}$

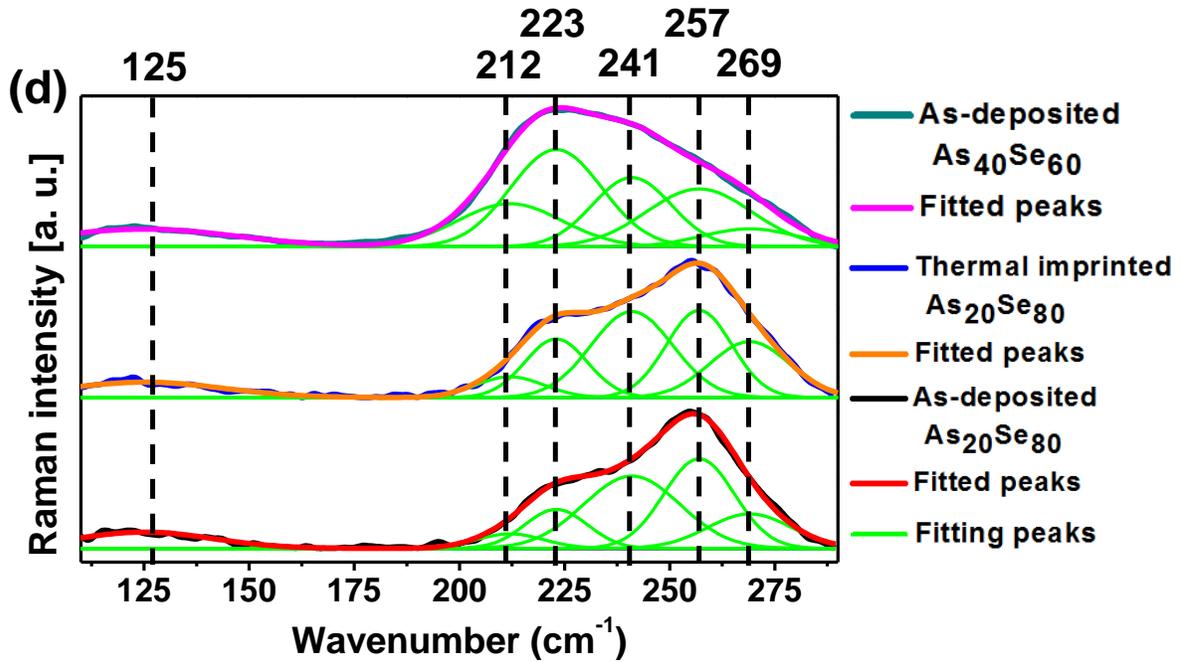

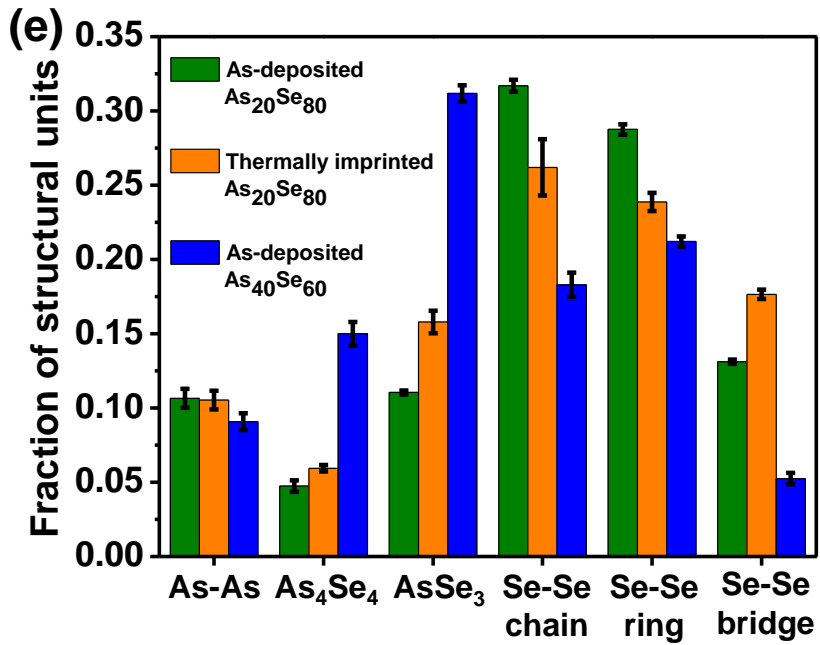

**Figure 2**. As-Se Chalcogenide Glass Film Characterizations (a) Cross-sectional SEM image of a thermally evaporated $As_{20}Se_{80}$ film on a $SiO_2$ coated silicon substrate. Thickness of the glass film is 360 nm. (b) Refractive index n and extinction coefficient k of $As_{40}Se_{60}$ and $Ss_{20}Se_{80}$ thin films measured by ellipsometer. (c) Raman spectra of as-deposited and thermally imprinted $As_{20}Se_{80}$

and As$_{40}$Se$_{60}$ thin films. (d) Raman spectra decomposition into the respective vibrational modes shown in **Table 1**. (e) The area fractions of each vibrational mode peak over the total area of all peaks, which are proportional to the concentration of each structural unit.

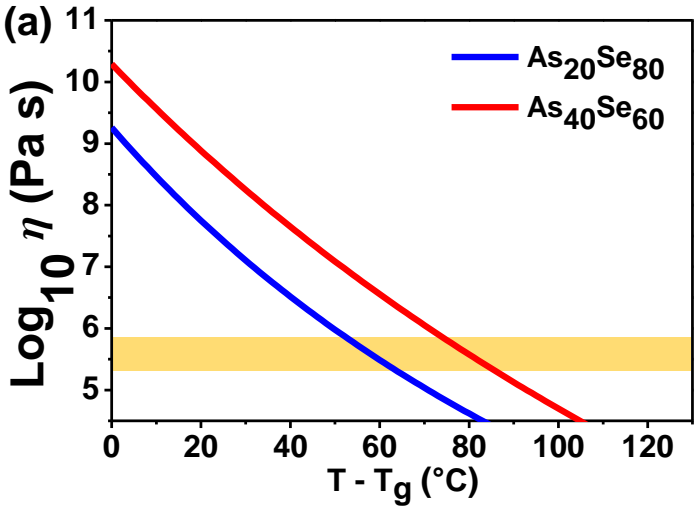

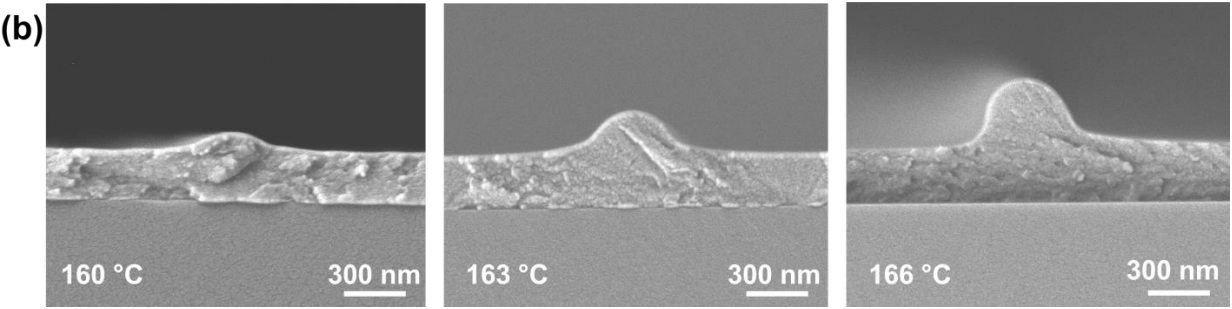

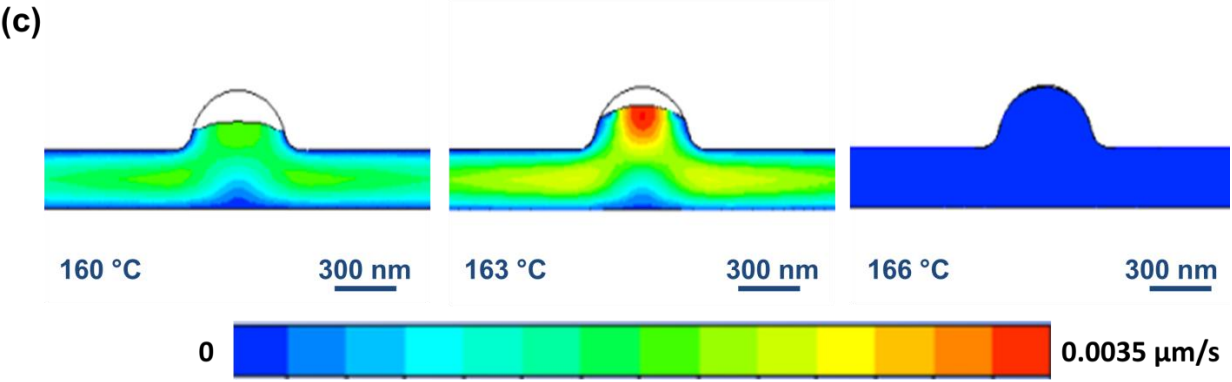

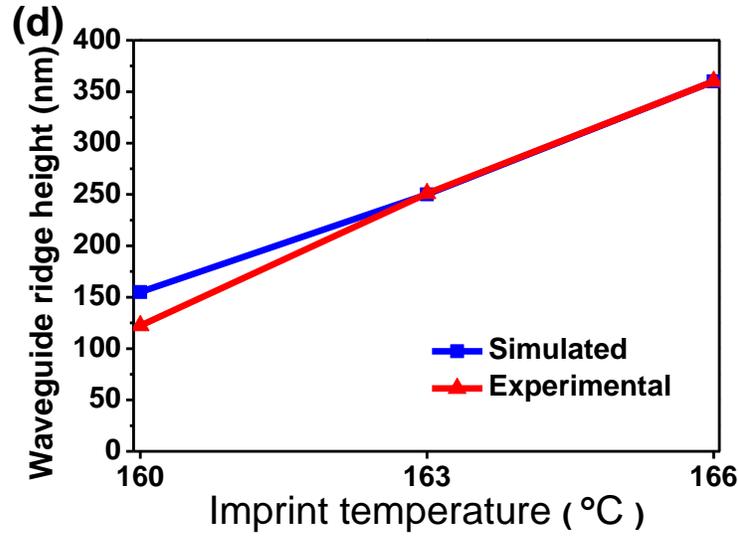

**Figure 3.** Kinetics of Geometry Shaping in Nanoimprint. (a) Viscosity of $As_{20}Se_{80}$ and $As_{40}Se_{60}$ (i.e. stoichiometric $As_{40}Se_{60}$) glasses plotted as functions of temperature based on the VFT model; the horizontal yellow stripe indicates the viscosity regime suitable for thermal nanoimprint fabrication; (b) cross-sectional morphology of imprinted ChG waveguides under SEM. The imprint temperatures are 160 ℃, 163 ℃ and 166 ℃, respectively; (c) cross-sectional geometry of imprinted waveguides calculated using finite element fluid dynamics simulations. The color scale indicates the velocity field of glass material. The VFT model was used to calculate the viscosity values at different temperatures: 160 ℃, 163 ℃ and 166 ℃ for the simulations, (d) a comparison of the imprinted ridge waveguide height between simulated prediction and experimental results, the experimental results were consistent with only 5 nm standard deviation.

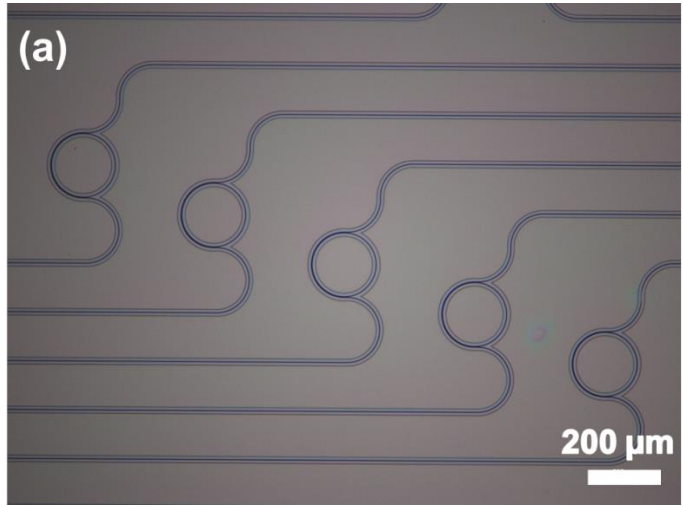

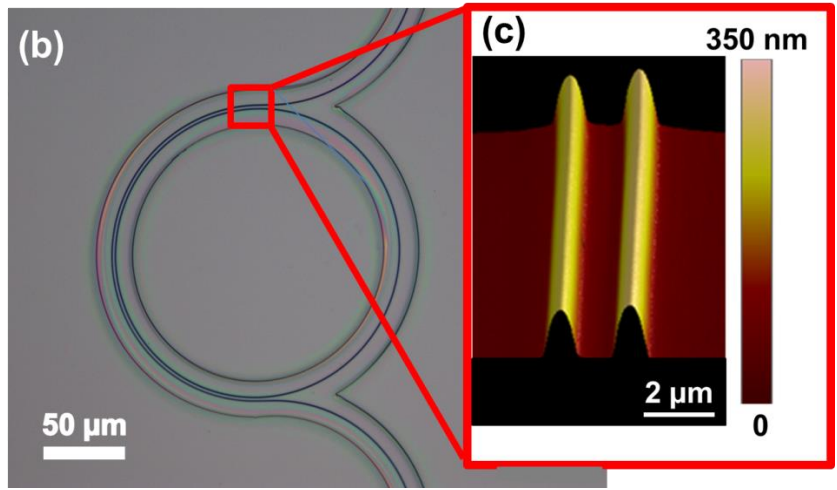

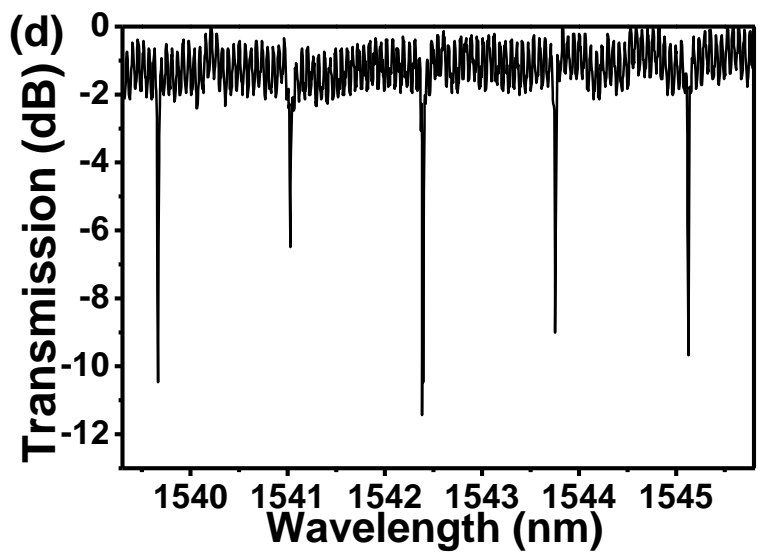

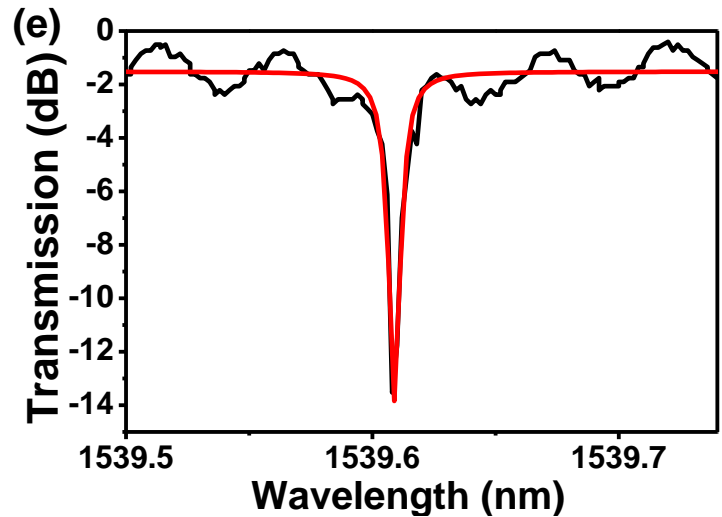

**Figure 4.** Optical Characterizations of Imprinted Devices. (a) and (b) Top-view microscopes image of imprinted $As_{20}Se_{80}$ micro-ring resonators with a radius of 100 µm; (c) surface morphology of a coupling section between a bus waveguide and a micro-ring measured by AFM, showing an RMS surface roughness of 0.8 nm. The coupling gap width is 800 nm; (d) TE-polarization transmission spectrum of an imprinted micro-ring resonator with an FSR of 1.09 nm; (e) transmission spectrum of a high-Q resonance showing an intrinsic Q-factor of 390,000, corresponding to an equivalent waveguide loss of 1.6 dB/cm.

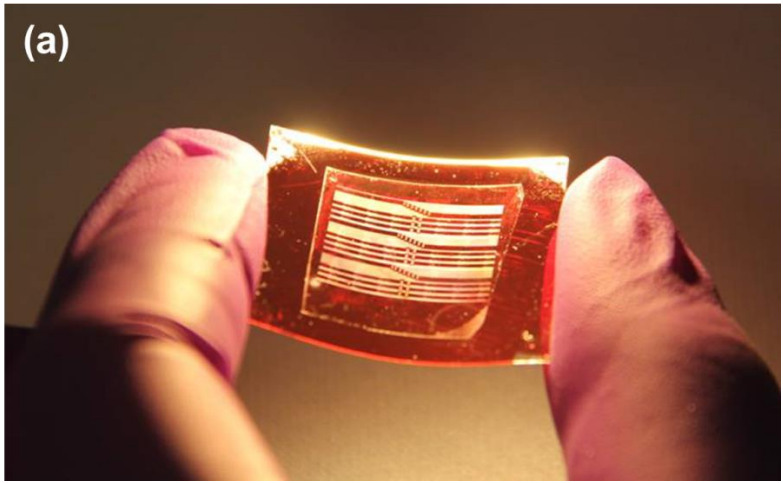

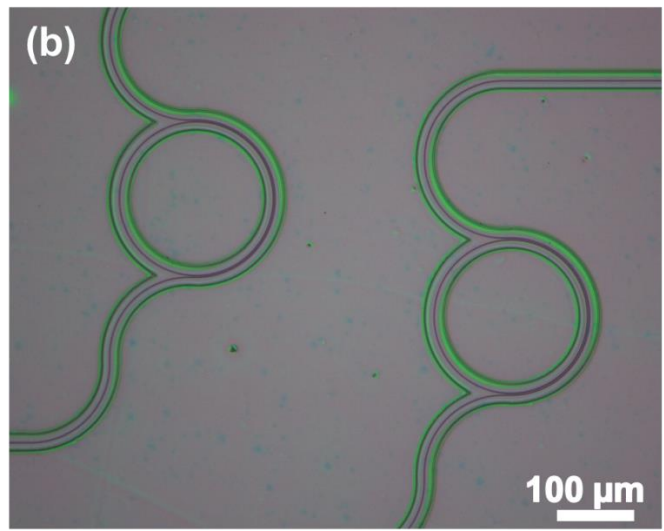

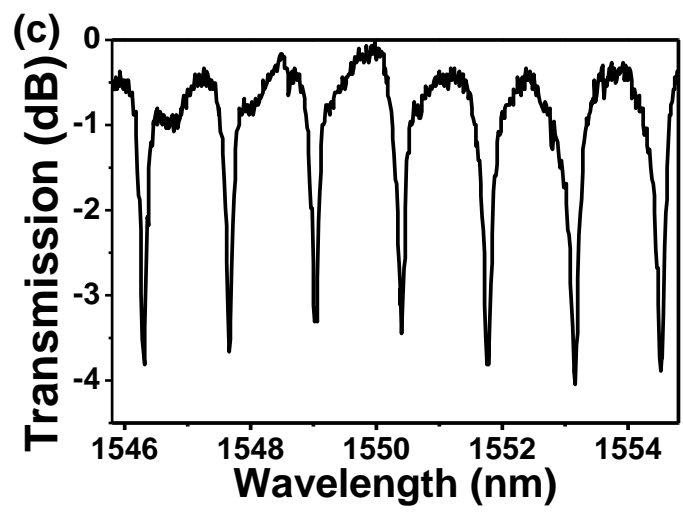

**Figure 5.** Direct Monolithic Imprint Fabrication on Nonplanar Substrates. (a) A photo showing imprinted ChG resonator devices on a flexible PET substrate; (b) top-view microscope image of ChG ring resonators imprinted on a flexible polyimide film; (c) transmission spectrum of an imprinted $As_{20}Se_{80}$ micro-ring resonator on a flexible substrate: an average Q = 10,000 was measured.

**Table 1.** Raman peak positions and assignments

| Peak Position (Wave number, $cm^{-1}$) | Peak assignment |
|---|---|
| 125 | As-As unit |
| 212 | $As_4Se_4$ unit |
| 223 | $AsSe_3$ pyramid unit |
| 241 | Se-Se chain |
| 257 | Se-Se ring |
| 269 | Se-Se bridge |